\documentclass[useAMS,usenatbib]{mn2e}
\usepackage{graphicx}
\def\rmscr#1{{\hbox{\rm \scriptsize #1}}}
\def\rmmat#1{{\hbox{\rm #1}}}
\def\d{\rmmat{d}}

\begin{document}
\title[White-Dwarf Kicks]{Constraining white-dwarf kicks in globular clusters}

\author[J. Heyl]{Jeremy Heyl$^{1}$\\
$^{1}$Department of Physics and Astronomy, University of British Columbia, Vancouver, British Columbia, Canada, V6T 1Z1 \\
Email: heyl@phas.ubc.ca; Canada Research Chair}

\date{\today}

\pagerange{\pageref{firstpage}--\pageref{lastpage}} \pubyear{2007}

\maketitle

\label{firstpage}

\begin{abstract}
  The wind of an asymptotic-giant-branch stars is sufficiently strong
  that if it is slightly asymmetric, it can propel the star outside of
  the open cluster of its birth or significantly alter its trajectory
  through a globular cluster; therefore, if these stellar winds are
  asymmetric, one would expect a deficit of white dwarfs of all ages
  in open clusters and for young white dwarfs to be less radially
  concentrated than either their progenitors or older white dwarfs in
  globular clusters.  This latter effect has recently been
  observed. Hence, detailed studies of the radial distribution of young white
  dwarfs in globular clusters could provide a unique probe of mass
  loss on the asymptotic giant branch and during the formation of
  planetary nebulae both as a function of metallicity and a limited
  range of stellar mass.
\end{abstract}

\section{Introduction}

\citet{1998A&A...333..603S} proposed that white dwarfs can acquire
their observed rotation rates from mild kicks generated by asymmetric
winds toward the end of their time on the asymptotic giant branch
(AGB) \citep{1993ApJ...413..641V}.  \citet{2003ApJ...595L..53F}
envoked these mild kicks to explain a putative dearth of white dwarfs
in open clusters
\citep[e.g.][]{1977A&A....59..411W,2001AJ....122.3239K}.
Unfortunately from the dynamics of open clusters it is difficult to
probe the kicks because the white dwarfs simply leave the cluster.
Although one can probe the dynamics of AGB winds directly through
observations of masers, attempts to look for asymmetries are dogged by
the variability of the star itself which makes it difficult to
constrain any relative motion between the centre of mass of the wind
and that of the star.  The dynamics of globular clusters provide a
unique environment to probe white dwarf kicks because both the
magnitude of the expected kick and the velocity dispersion of the
giants within the cluster are on the order  of kilometers per second,
so one would expect a significant signal.  On the other hand the
escape velocity of the cluster may be several times larger so most of
the white dwarfs remain in the cluster to measure after the kick.  The
expected signature of white dwarf kicks has been observed in M4 and
NGC~6397 \citep{2007Davis}.

This letter will examine how white dwarf kicks affect the radial
distribution of young white dwarfs in globular clusters through
analytic and Monte Carlo calculations of the phase-space distribution
function of stars in the cluster.

\section{Calculations}
Clusters of stars can typically be modelled with a lowered isothermal
profile (or King model)
\citep{1963MNRAS.125..127M,1966AJ.....71...64K}
\begin{equation}
f = \frac{\d N}{\d^3 x \d^3 v} = 
\left \{ 
\begin{array}{ll}
\rho_1(2\pi\sigma^2)^{-3/2} \left ( e^{\epsilon/\sigma^2}- 1 \right )
&  \rmmat{~if~} \epsilon>0 \\
0 & \rmmat{~if~} \epsilon\leq 0 
\end{array}
\right .
\label{eq:1}
\end{equation}
where $\epsilon = \Psi - \frac{1}{2} v^2$, $\Psi$ is the gravitational
potential, $\sigma$ is a characteristic velocity dispersion and
$\rho_1$ is a characteristic density.  Integrating the distribution
function over velocity yields the density distribution
\begin{equation}
\rho = \rho_1 \left [ e^{\Psi/\sigma^2} \rmmat{erf} \left ( \frac{\sqrt{\Psi}}{\sigma} \right ) + \sqrt{\frac{4\Psi}{\pi\sigma^2}} \left ( 1 + \frac{2\Psi}{3\sigma^2} \right )\right ].
\label{eq:2}
\end{equation}
The density is typically constant within the core radius,
$r_0=\sqrt{9\sigma^2/(4\pi\rho_0)}$ where $\rho_0$ is the central density of
the cluster.  The gravitational potential can be solved
self-consistently with Eq.~(\ref{eq:2}) \citep{Binn87} using
\begin{equation}
\frac{\d}{\d r} \left ( r^2 \frac{\d\Psi}{\d r} \right ) -4\pi G \rho r^2
\label{eq:3}
\end{equation}
to give a model for the cluster.
Because the distribution function depends only on constants of the
motion (the energy), it is constant in time as well.

With time the kinetic energy within the cluster approaches
equipartition between the various stars such that $m_i \sigma_i^2 =
m_j \sigma_j^2$ \citep{Spit87}.  The progenitors of young white dwarfs
will be the most massive main-sequence stars in a cluster at the time,
so they will typically have
$\sigma_\rmscr{TO}<\sigma_\rmscr{c}$, where
$\sigma_\rmscr{c}$ is the mean velocity dispersion of the
cluster; furthermore, because these progenitors will only have a small
fraction of the mass of the cluster, they can be considered as
massless tracers; their phase space density will be given by
Eq.~(\ref{eq:1}) with $\sigma=\sigma_\rmscr{TO}$.  The gravitational
potential and core radius will be determined by
Eq.~(\ref{eq:1})--Eq.~(\ref{eq:3}) with
$\sigma=\sigma_c$.

During their time on the AGB, the stars may lose a large fraction of
their mass suddenly and asymmetrically
\citep{1993ApJ...413..641V,1998A&A...333..603S,2003ApJ...595L..53F} and
receive an impulsive kick.  To model this kick, the original
phase-space density can be convolved with a kick distributed as a Gaussian:
\begin{equation}
f_\rmscr{final} = \frac{1}{\left (\pi \sigma_k^2\right)^{3/2}} \int \d^3 v' \exp \left [ - \frac{({\bf v} - {\bf v'})^2}{2 \sigma_k^2} \right ] f \left (x,v' \right ) .
\label{eq:4}
\end{equation}
Although this convolved distribution function can be expressed in
closed form (see the appendix), two key results are important.  First,
the final distribution function depends on the value of the
gravitational potential and the velocity separately and not on the
conserved energy alone.  Second, in the limit of large $\Psi$ or small
$\sigma_k$ the distribution approaches a lowered isothermal profile
with $\sigma_{\rm f}^2 \approx \sigma_\rmscr{TO}^2 + \sigma_k^2$.

Because the distribution no longer depends on constants of the motion
alone, the distribution function itself will depend on time, so a
Monte Carlo realization of the kicked distribution function is helpful
to make further progress.  Specifically, for a variety of values of
$\Psi(0)$, $\sigma_\rmscr{TO}$ and $\sigma_k$, ten thousand stars are
drawn from Eq.~(\ref{eq:1}) and given a three-dimensional velocity
kick drawn from a Gaussian of width of $\sigma_k$.  Finally, each star
is evolved forward along its orbit to a random phase; this yields an
estimate of the radial distribution of the young white dwarfs in the
cluster after the kick as well as the fraction of young white dwarfs
that might escape the cluster.  The potential of the cluster is fixed
to be the solution to Eq.~(\ref{eq:3}) for $\sigma=1$; this assumes
that the evolving AGB stars and young white dwarfs make a negligible
contribution to the mass of the cluster and that the observations of
the young white dwarfs occur within a relaxation time of their
formation.

\begin{figure}
\includegraphics[width=3.4in]{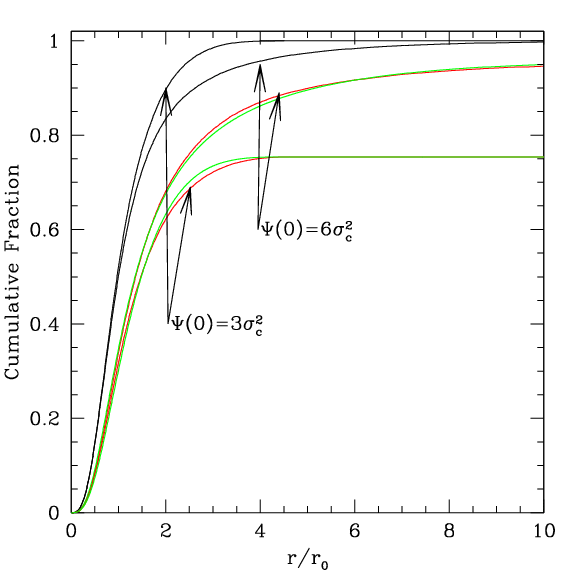}
\caption{The cumulative radial distribution for
  $\sigma_\rmscr{TO}=\sigma_k=0.7\sigma_c$ with $\Psi(0)=3\sigma_c^2$ and
  $6\sigma_c^2$.  The black curve follows the initial distribution, the
  red curve follows the phase-mixed distribution after the kick.  The
  green curves give the best fitting functions of the form
  Eq.~(\ref{eq:2}): a model with $\Psi(0)=3\sigma_c^2$ yields a escape
  fraction of 25\% and $\sigma_f=0.86\sigma_c$, and the model with
  $\Psi(0)=6\sigma_c^2$ gives an escape fraction of 4.4\% and
  $\sigma_f=0.76\sigma_c$. }
\label{fig:example}
\end{figure}

The result of the Monte Carlo realization is a list of radial
positions at the moment of the kick and at a random time later for
those stars that cannot escape the cluster.  The density distribution
given by Eq.~(\ref{eq:2}) provides a natural characterization of the
positions of the particles that do not manage to escape the cluster.
The best distribution is obtained by maximizing the Kolmogorov-Smirnov
probability \citep[e.g.][]{Pres92} that the list of radial positions
is drawn from Eq.~(\ref{eq:2}) with a particular value of $\sigma_f$;
in this way the escape fraction and the best-fit $\sigma_f$ provide an
approximation of the distribution of the young white dwarfs (see
Fig.~\ref{fig:example}).

\section{Results}
\label{sec:results}
\begin{figure}
\includegraphics[width=3.4in]{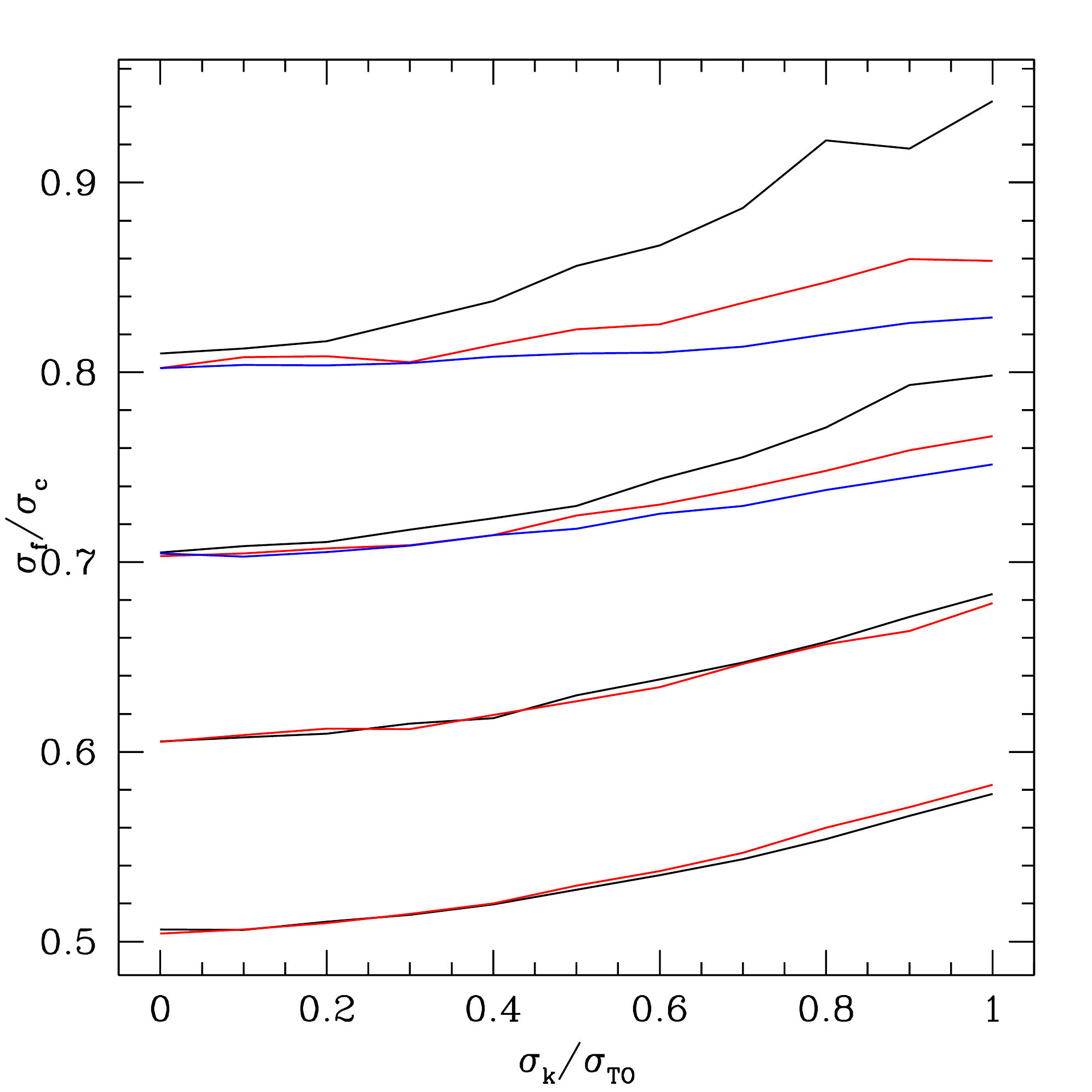}
\caption{The final values of the best fitting $\sigma$ for the
  phase-mixed stellar distributions as a function of the initial value
  of $\sigma_\rmscr{TO}/\sigma_c$, $\Psi(0)/\sigma_c^2$ and
  $\sigma_k/\sigma_\rmscr{TO}$.  From top to bottom the sets of
  curves are for $\sigma_\rmscr{TO}/\sigma_c=0.8$, 0.7, 0.6 and 0.5.
  Within each set the curves give
  $\Psi(0)/\sigma_c^2=4, 6$ and 8 in black, red and blue respectively.}
\label{fig:sigmaf}
\end{figure}

The numerical results depicted in Fig.~\ref{fig:sigmaf} confirm the
analytic expectations.  For small kicks the final distribution is well
characterized by 
\begin{equation}
\sigma_f^2\approx \sigma_\rmscr{TO}^2+\sigma_k^2
\label{eq:5}
\end{equation}
and in general $\sigma_f^2 < \sigma_\rmscr{TO}^2+\sigma_k^2$;
therefore, $\sigma_f^2-\sigma_\rmscr{TO}^2$, the estimate used by
\citet{2007Davis}, provides a firm lower limit on the kick velocity
required to puff up a stellar distribution within a cluster.  A kick
of a given size is less effective at changing the radial distribution
of stars within clusters with deeper potential wells than in shallow
wells.  

The second diagnostic is the fraction of the stars that escape the
cluster after receiving a kick.  The escape fraction is approximately
proportional to $\sigma_k^2/\sigma_\rmscr{TO}^2$, the relative change
in the kinetic energy of the stars due to the kick.  Furthermore, the
fraction decreases exponentially with increasing values of $\Psi(0)$
and increases exponetially with increasing values of the initial
velocity dispersion.  Lighter stars are much more likely to escape
from clusters with more shallow potential wells than heavier stars
that are initially more centrally concentrated.  More importantly and
in contrast with the results for open clusters
\citep{2003ApJ...595L..53F}, the escape fraction is typically quite
small; it did not exceed 25\% for any of the models considered (not
surprisingly the largest number of escapees came from a model with
$\Phi(0)=4\sigma_c^2$ and $\sigma_k=\sigma_\rmscr{TO}=0.8\sigma_c$).
\begin{figure}
\includegraphics[width=3.4in]{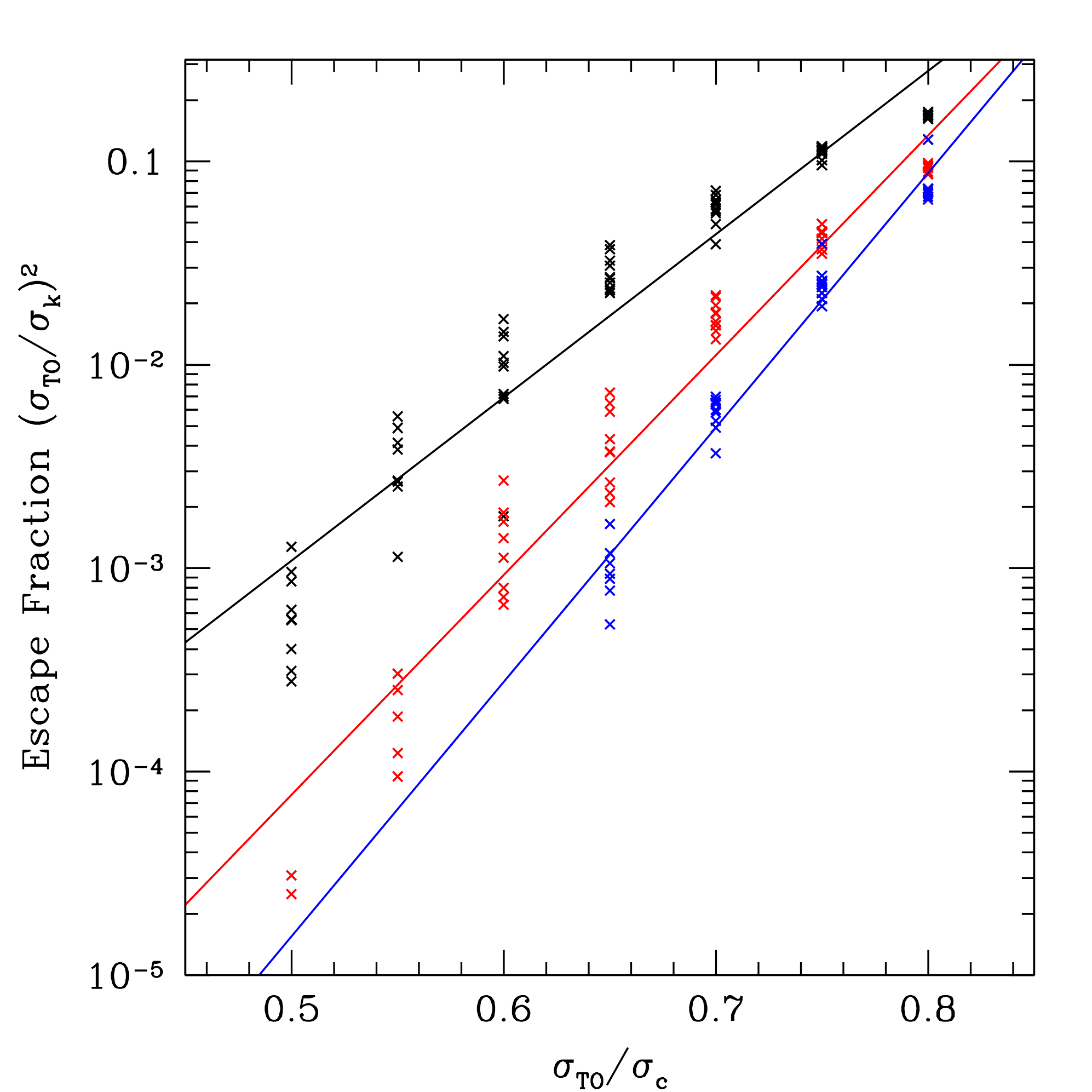}
\caption{The fraction of stars that escape the cluster after receiving
  a kick.  The escape fraction is typically proportional to the
  square of the relative size of the kick, so this dependence has
  been removed.  The black, red and blue points give the results for 
  $\Psi(0)/\sigma_c^2=4, 6$ and 8, respectively, for various values
  of $\sigma_k$ ranging from $0.1 \sigma_\rmscr{TO}$ to $\sigma_\rmscr{TO}$, and
  values of $\sigma_\rmscr{TO}$ from 0.5 to 0.8$\sigma_c$.
  Where the points are absent the 
  escape fraction was less than one part in $10^4$.
}
\label{fig:escape}
\end{figure}

Depending on the distribution of stellar masses in the cluster the
ratio of $\sigma_\rmscr{TO}$ to $\sigma_c$ may vary.  For simplicity,
the mass function can be assumed to be a power law,
\begin{equation}
\frac{dN}{dM} = A M^{-\alpha} ~\rmmat{for}~ M_\rmscr{min} \leq M \leq
M_\rmscr{TO}
\label{eq:6}
\end{equation}
and zero elsewhere.  This yields a mean mass of
%
%
%
\begin{equation}
{\bar M} = \frac{1-\alpha}{2-\alpha} M_\rmscr{TO} \frac{
  \left(\frac{M_\rmscr{min}}{M_\rmscr{TO}} \right )^{2-\alpha}-1}{
\left(\frac{M_\rmscr{min}}{M_\rmscr{TO}} \right )^{1-\alpha}-1}.
\label{eq:7}
\end{equation}
If $M_\rmscr{min}=0.1 M_\rmscr{TO}$ and $\alpha=-2.25$ (a Salpeter
initial mass function), $\bar M \approx 0.23 M_\rmscr{TO}$ and
$\sigma_{TO}=0.5\sigma_c$.

Typically when the line-of-sight velocity dispersion of a cluster is
determined, the result is dominated by the brightest stars, that is
the giants and the stars near the turnoff; therefore, the
line-of-sight velocity dispersion measured near the centre of the
cluster is a direct determination of the value of
$\sigma_\rmscr{TO}$. Typically, $\sigma_\rmscr{los} \rightarrow
\sigma_\rmscr{TO}$ as $\Psi(0) \rightarrow \infty$.  Specifically,
\begin{equation}
\sigma_\rmscr{los} \approx \sigma_\rmscr{TO} \left [ 1 -
  \exp \left (-\frac{\left(\Psi(0)/\sigma_\rmscr{TO}^2\right)^{0.85}}{1.45} \right ) \right ]
\label{eq:8}
\end{equation}
to within two percent for $\Psi(0)>2\sigma_\rmscr{TO}^2$.  For
$\Psi(0)=4\sigma_\rmscr{TO}^2$, $\sigma_\rmscr{los}=0.89\sigma$.  By
looking at the distribution of stellar masses in the cluster, the
value of $\sigma_c$ can be estimated from $\sigma_\rmscr{TO}$ as in
Eq.~(\ref{eq:6})~and~(\ref{eq:7}).  As Tab.~\ref{tab:gc} shows, the
line-of-sight velocity dispersion of a globular cluster is typically a
few kilometers per second, similar to the proposed kick velocities;
the two clusters observed by \citet{2007Davis} have velocity
dispersions of about 4~kms$^{-1}$, possibly smaller than the expected
kicks.  Furthermore, the effect could be observable in all of the
clusters listed with the possible exceptions of 47 Tuc, Omega Cen and M22.
\begin{table*}
  \caption{Kinetic parameters for a few nearby globular clusters whose
    radial stellar distribution could be probed with JWST (Richer {\em
      priv. comm.}).  
    The concentration or the ratio of the core radius
    to the tidal radius ($c=\log_{10} r_t/r_c$) are from \citet{Harr96}, a
    dash denotes that the core has collapsed.
    The values of $\Psi(0)/\sigma_c^2$ are from
    \citet{2005ApJS..161..304M} with the exception of NGC 6837 where
    it was
    inferred from the concentration.   The value of
    $\sigma_c$ was obtained by assuming that ${\bar
      M}/M_\rmscr{TO}=0.3/0.8=0.375$. Refs: (a) \citet{2006ApJS..166..249M}; (b)
    \citet{2006A&A...445..503R}; (c) \citet{1986ApJ...305..645P}; (d) \citet{1993ASPC...50..357P}. }
\label{tab:gc}
\begin{center}
  \begin{tabular}{l|rrrr}
    \hline
\hline
\multicolumn{1}{c}{Cluster} &
\multicolumn{1}{c}{$\sigma_\rmscr{los}$ [km/s]} &
\multicolumn{1}{c}{$\sigma_c$  [km/s]} &
\multicolumn{1}{c}{$c$} &
\multicolumn{1}{c}{$\Psi(0)/\sigma_c^2$} \\
\hline 
NGC 104 (47 Tuc) & 11.6$\pm$0.8$^\rmscr{(a)}$ & 19 & 2.03 & 8.60$\pm$0.10 \\
NGC 5139 (Omega Cen) & 15$^\rmscr{(b)}$ & 25 & 1.61 & 6.20$\pm$0.20 \\
NGC 6121 (M4) & 3.9$\pm$0.7$^\rmscr{(c)}$  & 6.4 & 1.59 & 7.40$\pm$0.10 \\
NGC 6397 & 3.5$\pm$0.2$^\rmscr{(d)}$ & 5.7 &  \multicolumn{1}{c}{---}  & \multicolumn{1}{c}{---} \\
NGC 6656 (M22) &  8.5$\pm$1.9$^\rmscr{(c)}$ & 14 & 1.31 & 6.50$\pm$0.20 \\
NGC 6752 & 4.5$\pm$0.5$^\rmscr{(d)}$ & 7.3 & \multicolumn{1}{c}{---} &\multicolumn{1}{c}{---} \\
NGC 6809 (M55) & 4.2$\pm$0.5$^\rmscr{(d)}$ & 6.8 & 0.76 & 4.50$\pm$0.10 \\
NGC 6838 (M71)&  2.8$\pm$0.6$^\rmscr{(c)}$ & 4.6 &  1.15 & 5.4 \\
\hline
  \end{tabular}
\end{center}
\end{table*}

\section{Conclusions}

Studies of the radial distribution of stars near the turnoff and young
white dwarfs within globular clusters can provide a unique handle on
the asymmetry of the rapid mass loss while stars are on the asymptotic
giant branch.  In the case of an open cluster as studied by
\citet{2003ApJ...595L..53F}, the assumed dearth of young white dwarfs
could only provide an lower limit on the kick; to get a better result
requires searching for the white dwarfs that have just left the
cluster.  In a globular cluster one can get a quantitative estimate of
the typical kick velocity from the observed distributions because very
few stars typically escape the cluster; furthermore, by comparing the
distributions in globular clusters of different ages and metallicities
one can probe the mass loss for a variety of stellar masses and
chemical compositions.  By comparing the distribution of white dwarfs
of different ages up to and greater than the relaxation time for the
cluster, the process of dynamical relaxation of the white dwarf
population can be constrained observationally.

Such a study requires observations of many stars throughout a globular
cluster, specifically the accurate radial positions of the brighest
main-sequence stars (near the turnoff) and the brightest white dwarfs.
Because information from the central regions of the cluster is most
helpful, understanding or at least minimizing the effects of confusion
is crucial; the space-based observations presented in
\citep{2007Davis} indeed found evidence of white dwarf kicks in NGC~6397
and M4.  Further studies of resolved stellar populations in both open
clusters and globular clusters could provide further probes of this
violent stage in the evolution of stars like our Sun.

\section*{Appendix : Kicked Phase-Space Distribution Function}
\label{sec:appendix-:-kicked}

The convolution of the lowered isothermal profile distribution
function with a Gaussian kick in velocity is given by
\begin{equation}
f_\rmscr{final} = \frac{1}{\left (\pi \sigma_k^2\right)^{3/2}} \int \d^3 v' \exp \left [ - \frac{({\bf v} - {\bf v'})^2}{2 \sigma_k^2} \right ] f_\rmscr{initial}\left (x,v' \right ) .
\end{equation}
Because both the kick and basic distribution function are Gaussian it is
not surprising that the final distribution function can be expressed in
closed form.  If there was no restriction on the velocities in the
distribution function in Eq.~(\ref{eq:1}), the convolution would also
be a Gaussian with
$\sigma^2_f = \sigma^2+\sigma^2_k$; however, the
restriction on the velocities complicates the result:
\begin{eqnarray}
f_\rmscr{final} = N \Biggr \{ 
\frac{2 \sigma_k \sigma_f}{v} e^{-(\Psi+v^2/2)/\sigma_k^2} \left ( 1 - e^{-2\sqrt{2\Psi} v/\sigma_k^2} \right )  + \nonumber \\
~~~~~ \sqrt{2\pi} \sigma \exp \left [ \frac{\epsilon}{\sigma_f^2} + \frac{\sigma_k \Psi}{\sigma^2 \sigma_f^2} + \frac{v \sqrt{2\Psi}}{\sigma_k^2} \right ] \left [ g(v) + g(-v) \right ] \Biggr \}
\end{eqnarray}
where
\begin{eqnarray}
N &=&  \frac{\rho_1}{4 \sqrt{2} \pi^2 \sigma \sigma_f^3} \exp \left [ - \frac{v\sqrt{2\Psi}}{\sigma_k^2} - 1 \right ] \\
g(\Psi,v) &=& \rmmat{erf} \left ( \frac{\sigma^2 v + \sigma_f^2 \sqrt{2\Psi}}{\sqrt{2} \sigma \sigma_k \sigma_f} \right ) 
\end{eqnarray}
In the limit of large $\Psi$ or small $\sigma_k$ the distribution
approaches a lowered isothermal profile with $\sigma\rightarrow
\sigma_f$.
 
\section*{Acknowledgments}
I would like to thank Harvey Richer and Saul Davis for useful
discussions.  The Natural Sciences and Engineering Research Council of
Canada, Canadian Foundation for Innovation and the British Columbia
Knowledge Development Fund supported this work.  Correspondence and
requests for materials should be addressed to heyl@phas.ubc.ca.  This
research has made use of NASA's Astrophysics Data System Bibliographic
Services

\bibliographystyle{mn2e}
\bibliography{mine,wd,physics,math}

\begin{thebibliography}{}

\bibitem[\protect\citeauthoryear{Binney \& Tremaine}{Binney \&
  Tremaine}{1987}]{Binn87}
Binney J.,  Tremaine S.,  1987, Galactic Dynamics.
Princeton Univ. Press, Princeton

\bibitem[\protect\citeauthoryear{{Davis}, {Richer}, {Coffey}, {Anderson},
  {Brewer}, {Fahlman}, {Hansen}, {Hurley}, {Kalirai}, {King}, {Reitzel},
  {Rich}, {Rich} \& {Shara}}{{Davis} et~al.}{2006}]{2007Davis}
{Davis} S.,  {Richer} H.~B.,  {Coffey} J.,  {Anderson} J.,  {Brewer} J.,
  {Fahlman} G.~G.,  {Hansen} B.~M.,  {Hurley} J.,  {Kalirai} J.~S.,  {King}
  I.~R.,  {Reitzel} D.,  {Rich} R.~M.,  {Rich} M.~R.,    {Shara} M.~M.,  2006,
  in American Astronomical Society Meeting Abstracts Vol.~209 of American
  Astronomical Society Meeting Abstracts, {Are white dwarfs born with a
  `KICK'?}.
pp \#228.03--+

\bibitem[\protect\citeauthoryear{{Fellhauer}, {Lin}, {Bolte}, {Aarseth} \&
  {Williams}}{{Fellhauer} et~al.}{2003}]{2003ApJ...595L..53F}
{Fellhauer} M.,  {Lin} D.~N.~C.,  {Bolte} M.,  {Aarseth} S.~J.,    {Williams}
  K.~A.,  2003, \apjl, 595, L53

\bibitem[\protect\citeauthoryear{Harris}{Harris}{1996}]{Harr96}
Harris W.~E.,  1996, \aj, 112, 1487

\bibitem[\protect\citeauthoryear{{Kalirai}, {Ventura}, {Richer}, {Fahlman},
  {Durrell}, {D'Antona} \& {Marconi}}{{Kalirai}
  et~al.}{2001}]{2001AJ....122.3239K}
{Kalirai} J.~S.,  {Ventura} P.,  {Richer} H.~B.,  {Fahlman} G.~G.,  {Durrell}
  P.~R.,  {D'Antona} F.,    {Marconi} G.,  2001, \aj, 122, 3239

\bibitem[\protect\citeauthoryear{{King}}{{King}}{1966}]{1966AJ.....71...64K}
{King} I.~R.,  1966, \aj, 71, 64

\bibitem[\protect\citeauthoryear{{McLaughlin}, {Anderson}, {Meylan},
  {Gebhardt}, {Pryor}, {Minniti} \& {Phinney}}{{McLaughlin}
  et~al.}{2006}]{2006ApJS..166..249M}
{McLaughlin} D.~E.,  {Anderson} J.,  {Meylan} G.,  {Gebhardt} K.,  {Pryor} C.,
  {Minniti} D.,    {Phinney} S.,  2006, \apjs, 166, 249

\bibitem[\protect\citeauthoryear{{McLaughlin} \& {van der Marel}}{{McLaughlin}
  \& {van der Marel}}{2005}]{2005ApJS..161..304M}
{McLaughlin} D.~E.,  {van der Marel} R.~P.,  2005, \apjs, 161, 304

\bibitem[\protect\citeauthoryear{{Michie}}{{Michie}}{1963}]{1963MNRAS.125..127%
M}
{Michie} R.~W.,  1963, \mnras, 125, 127

\bibitem[\protect\citeauthoryear{{Peterson} \& {Latham}}{{Peterson} \&
  {Latham}}{1986}]{1986ApJ...305..645P}
{Peterson} R.~C.,  {Latham} D.~W.,  1986, \apj, 305, 645

\bibitem[\protect\citeauthoryear{Press, Teukolsky, Vettering \& Flannery}{Press
  et~al.}{1992}]{Pres92}
Press W.~H.,  Teukolsky S.~A.,  Vettering W.~T.,    Flannery B.~P.,  1992,
  Numerical Recipes in C, second edn.
Cambridge Univ. Press, Cambridge

\bibitem[\protect\citeauthoryear{{Pryor} \& {Meylan}}{{Pryor} \&
  {Meylan}}{1993}]{1993ASPC...50..357P}
{Pryor} C.,  {Meylan} G.,  1993, in {Djorgovski} S.~G.,  {Meylan} G.,  eds,
  Structure and Dynamics of Globular Clusters Vol.~50 of Astronomical Society
  of the Pacific Conference Series, {Velocity Dispersions for Galactic Globular
  Clusters}.
pp 357--+

\bibitem[\protect\citeauthoryear{{Reijns}, {Seitzer}, {Arnold}, {Freeman},
  {Ingerson}, {van den Bosch}, {van de Ven} \& {de Zeeuw}}{{Reijns}
  et~al.}{2006}]{2006A&A...445..503R}
{Reijns} R.~A.,  {Seitzer} P.,  {Arnold} R.,  {Freeman} K.~C.,  {Ingerson} T.,
  {van den Bosch} R.~C.~E.,  {van de Ven} G.,    {de Zeeuw} P.~T.,  2006, \aap,
  445, 503

\bibitem[\protect\citeauthoryear{Spitzer}{Spitzer}{1987}]{Spit87}
Spitzer L.,  1987, Dynamical Evolution of Globular Clusters.
Princeton Univ. Press, Princeton

\bibitem[\protect\citeauthoryear{{Spruit}}{{Spruit}}{1998}]{1998A&A...333..603%
S}
{Spruit} H.~C.,  1998, \aap, 333, 603

\bibitem[\protect\citeauthoryear{{Vassiliadis} \& {Wood}}{{Vassiliadis} \&
  {Wood}}{1993}]{1993ApJ...413..641V}
{Vassiliadis} E.,  {Wood} P.~R.,  1993, \apj, 413, 641

\bibitem[\protect\citeauthoryear{{Weidemann}}{{Weidemann}}{1977}]{1977A&A....5%
9..411W}
{Weidemann} V.,  1977, \aap, 59, 411

\end{thebibliography}
\label{lastpage}
\end{document}